\documentclass[aps,showpacs,twocolumn] {revtex4}
\usepackage{graphicx}
\usepackage{bm}
\usepackage{amsfonts,amsmath,amssymb}
\usepackage[squaren]{SIunits}
\makeatletter
\begin{document}

\title{Transport properties and magnetic field induced localization in
the misfit cobaltite
[Bi$_2$Ba$_{1.3}$K$_{0.6}$Co$_{0.1}$]$^{RS}$[CoO$_2$]$_{1.97}$
single crystal}
\author{X. G. Luo, H. Chen, G. Y. Wang, G. Wu, T. Wu, L. Zhao, and X. H. Chen$^\ast$} \affiliation{Hefei National
Laboratory for Physical Science at Microscale and Department of
Physics, University of Science and Technology of China, Hefei, Anhui
230026, People's Republic of China}

\begin{abstract}
Resistivity under magnetic field, thermopower and Hall coefficient
are systematically studied for
[Bi$_2$Ba$_{1.3}$K$_{0.6}$Co$_{0.1}$]$^{RS}$[CoO$_2$]$_{1.97}$
single crystal. In-plane resistivity ($\rho_{ab}$(T)) shows
metallic behavior down to 2 K with a $T^2$ dependence below 30 K;
while out-of-plane resistivity ($\rho_{c}(T)$) shows metallic
behavior at high temperature and a thermal activation
semiconducting behavior below about 12 K. Striking feature is that
magnetic field induces a ln(1/$T$) diverging behavior in both
$\rho_{ab}$ and $\rho_{c}(T)$ at low temperature. The positive
magnetoresistance (MR) could be well fitted by the formula based
on multi-band electronic structure. The ln(1/$T$) diverging
behavior in $\rho_{ab}$ and $\rho_{c}(T)$ could arise from the
magnetic-field-induced 2D weak localization or spin density wave.
\end{abstract}


\maketitle

\section{INTRODUCTION}

Triangular cobaltites have attracted significant interest for the
promising application prospect as thermoelectrical materials and
the complex physical properties as strongly correlated electron
system.\cite{Teresaka,Maignan,WangYY,LiSY} The complex physical
properties include the unconventional superconductivity in
water-intercalated Na$_{0.35}$CoO$_{2}$,\cite{Takada}
temperature-dependent Hall effect,\cite{Yamamoto,WangYY} large
negative MR in (Bi,Pb)$_2M_2$Co$_2$O$_y$ (M=Sr and Ca) and
Ca$_3$Co$_4$O$_9$,\cite{Yamamoto,Maignan,Maignan2} large
thermopower (TEP) with low resistivity,\cite{Teresaka,Maignan} and
complicated magnetic structure,\cite{Cava,WangCH,Sugiyama} etc.
These triangular cobaltites have the common structural unit of
CdI$_2$-type hexagonal [CoO$_2$] layer, which is composed of
edge-shared CoO$_6$ octahedra. Among them, the so-called misfit
cobaltites Bi$_2M_2$Co$_2$O$_y$ ($M$=Ca, Sr and Ba)
\cite{Leigylin} and Ca$_3$Co$_4$O$_9$ \cite{Maignan} are
constructed by the alternative stacking rocksalt(RS)-type blocks
and [CoO$_2$] layers. The two sublattices share the common $a$-
and $c$-lattice parameters, but possess the different $b$-lattice
lengths causing a misfit along $b$-axis with a misfit ratio
$b_{RS}$/$b_H$ ($b_{RS}$ and $b_H$ are the $b$-lattice parameters
for RS and hexagonal sublattices, respectively). In
Bi$_2M_2$Co$_2$O$_y$ ($M$=Ca, Sr and Ba), the quadruple RS block
is composed of two deficient [BiO] layers sandwiched by two [$M$O]
layers.\cite{Leigylin} As $M$=Ba, commensurate modulation along
b-axis between RS and hexagonal sublattices with
$b_{RS}$/$b_H$=2.0 was found,\cite{Maignan3} being in contrast to
the incommensurate modulation in other misfit cobaltites.
Metallicity increases with increasing the ionic radii from Ca to
Ba,\cite{Maignan2,Yamamoto,Maignan3} and the MR at low temperature
also changes the sign from negative in Ca and Sr
compounds\cite{Maignan2,Yamamoto} to positive in Ba
compounds.\cite{Maignan3} It has been found by us that there
coexist large negative and positive contributions to the
nonmonotonic magnetic-field dependent MR in Pb-doped
Bi$_2$Sr$_2$Co$_2$O$_y$.\cite{Luo} Spin-dependent charge transport
has been taken into account for understanding the large negative
MR in Sr and Ca compounds, but the large positive MR in Ba
compounds has not been fully understood. In polycrystalline
Bi$_2$Ba$_2$Co$_2$O$_y$ sample, Hervieu et al. reported that the
positive isothermal MR exhibits a linear $H$ dependence. They
compared the behavior in Ba compounds with MR in Sr$_2$RuO$_4$ and
heavy-Fermion-like oxide LiV$_2$O$_4$ to understand the positive
MR.\cite{Maignan3} However, the origin of large positive MR was
not settled down,  further work, especially on single crystals, is
required to understand this anomalous positive MR.

In this article, K-doped Bi$_2$Ba$_2$Co$_2$O$_y$ single crystal
was grown through flux method. The single crystal shows metallic
resistivity in $ab$ plane down to 2 K, while exhibits a weak
thermal activation behavior along $c$-axis at low temperature.
Magnetic field induces a ln(1/$T$) diverging behavior at low
temperature in $\rho_{ab}(T)$ and $\rho_c(T)$.  The positive MR
can be interpreted by the multiband electronic structure. The
ln(1/$T$) diverging behavior in $\rho_{ab}(T)$ may be ascribed to
the magnetic-field-induced 2D weak localization or spin density
wave (SDW).

\section{EXPERIMENTAL PROCEDURES}
Bi-Ba-Co-O single crystals were grown by the solution method using
K$_2$CO$_3$-KCl fluxes. Starting materials Bi$_2$O$_3$, BaCO$_3$ and
Co$_3$O$_4$ were mixed in a proportion of Bi:Ba:Co = 2:2:2 with a
total weigh to be 4 grams. The powders were heated at 800$\celsius$
for 10 hours. Then the prepared Bi$_2$Ba$_2$Co$_2$O$_y$ was mixed
with the mixture of KCl and K$_2$CO$_3$ with a molar proportion of
1:4 (20.5 grams), which was loaded in an aluminum crucible having 30
ml volume. The solute concentration was about 1.5 mol$\%$. A lidded
crucible was used to prevent the solution from evaporating and to
grow crystals under stable conditions. The mixture was melted at
950$\celsius$ for 20 hours, and then slowly cooled down to
600$\celsius$ at a rate of 5$\celsius$/hr. The single crystals were
separated from the melt by washing with distilled water. The
crystals were large thin platelets and black in color. Typical
dimensions of the crystals are 5$\times$5$\times$0.05 mm$^3$.

Single crystals were characterized by electron diffraction (ED) and
x-ray diffractions (XRD) using Cu $K_{\alpha}$ radiations,
respectively. The actual chemical composition of the single crystals
was determined by inductively coupled plasma (ICP) atomic emission
spectroscopy (AES) (ICP-AES) technique and X-ray Energy Dispersive
Spectrum (EDS). The obtained results from ICP-AES and EDS were
almost consistent to be Bi: Ba: K: Co = 2: 1.3: 0.6: 2.1.

Electrical transport was measured using the ac four-probe method
with an alternative current (ac) resistance bridge system (Linear
Research, Inc.; LR-700P). Hall effect is measured by four-terminal
ac technique. To eliminate the offset voltage due to the asymmetric
Hall terminals, the magnetic field was changed from -5 to 5 T and
the Hall voltage was calculated to be $\{$$V(H)-V(-H)$$\}$/2, where
$V$ is the voltage between the Hall probes. The dc magnetic field
for MR measurements is supplied by a superconducting magnet system
(Oxford Instruments). Thermoelectric power (TEP) was measured using
the steady-state technique.

\section{EXPERIMENTAL RESULTS}

\begin{figure}[htbp]
\includegraphics[width=0.48\textwidth]{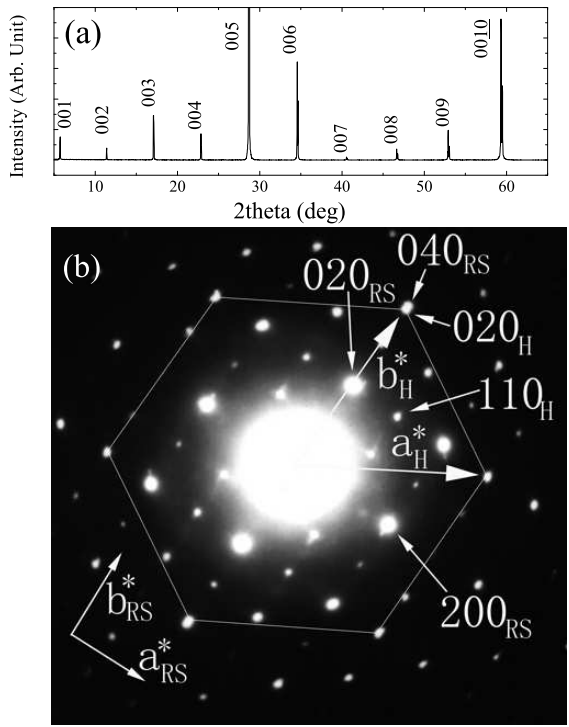}
\caption{(a):X-ray diffraction pattern; (b): [001] electron
diffraction pattern for the
[Bi$_2$Ba$_{1.3}$K$_{0.6}$Co$_{0.1}$]$^{RS}$[CoO$_2$]$_{1.97}$
single crystal}
\end{figure}

\subsection{Structural characterization}

XRD pattern shown in Fig. 1a indicates that the single crystals
are perfect c-orientation with the c-axis lattice parameter of
15.55 ${\rm \AA}$, larger than that in undoped polycrystalline
sample (15.44 ${\rm \AA}$),\cite{Maignan3} being consistent with
the substitution of larger K$^+$ for Ba$^{2+}$ ion
($r_{K^+_{VI}}$=1.38 ${\rm \AA}$, $r_{Ba^{2+}_{VI}}$=1.36 ${\rm
\AA}$).\cite{Shannon} The [001] ED pattern shown in Fig.1b is
similar to that of undoped sample reported by Hervieu et
al.\cite{Maignan3} except for that the reflection of (020)$_{H}$
visibly separates from that of (040)$_{RS}$ (where H and RS refer
to the hexagonal and RS sublattices, respectively). In-plane
lattice parameters can be estimated from the [001] ED pattern.
\textbf{a} and \textbf{b} parameter of RS sublattice is larger
than those in undoped polycrystalline sample. $a_{RS}$ and
$b_{RS}$ are 4.905 and 5.640 $\AA$ for undoped polycrystalline
sample, while 5.031 and 5.683 $\AA$ for the present single
crystal, respectively. This is due to the substitution of larger
K$^+$ for Ba$^{2+}$ ion. Along \textbf{a} axis, it is obtained
$a_{RS}$=$\sqrt{3}$$a_{H}$ ($a_{H}$=2.907$\AA$), indicating that
the RS and H subsystems share the common a-axis lattice parameter.
Along \textbf{b} axis, however, an incommensurate modulation with
the misfit ratio $b_{RS}/b_{H}$ = 1.97 ($b_{H}$=2.88$\AA$) can be
obtained, indicating that \textbf{b}$_{RS}$ and \textbf{b}$_{H}$
axes are colinear but aperiodic. This is in contrast to the
commensurate modulation along \textbf{b} direction in undoped
polycrystalline sample ($b_{RS}/b_{H}$ = 2).\cite{Maignan3}
Therefore, the structural formula of the present compound could be
written as
[Bi$_2$Ba$_{1.3}$K$_{0.6}$Co$_{0.1}$]$^{RS}$[CoO$_2$]$_{1.97}$.

\begin{figure}[htbp]
\includegraphics[width=0.45\textwidth]{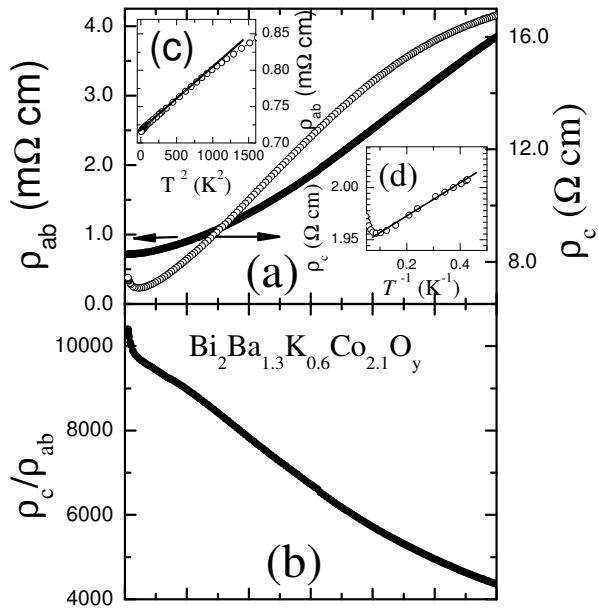}
\caption{ (a): Temperature dependence of in-plane and out-of-plane
resistivity; (b): Temperature dependence if resistivity anisotropy
$\rho_{c}/\rho_{c}$ for the
[Bi$_2$Ba$_{1.3}$K$_{0.6}$Co$_{0.1}$]$^{RS}$[CoO$_2$]$_{1.97}$
single crystal. (c): $T^2$ dependence of the in-plane resistivity
below 50 K; (d): Plot of ln($\rho_{c}$ vs (1/$T$)}
\end{figure}

\begin{figure}[htbp]
\includegraphics[width=0.48\textwidth]{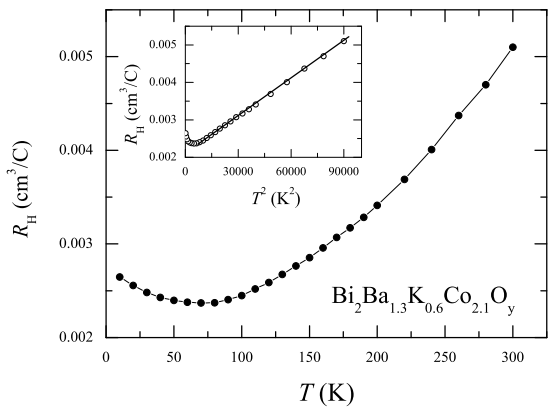}
\caption{Temperature dependence of the Hall coefficient for the
[Bi$_2$Ba$_{1.3}$K$_{0.6}$Co$_{0.1}$]$^{RS}$[CoO$_2$]$_{1.97}$
single crystal. Inset shows the square $T$ dependence of the Hall
coefficient.}
\end{figure}

\subsection{Resistivity, Hall coefficient and thermopower}

Figure 2a shows the temperature dependence of the in-plane and
out-of-plane resistivity ($\rho_{ab}$($T$) and $\rho_{c}$($T$))
for [Bi$_2$Ba$_{1.3}$K$_{0.6}$Co$_{0.1}$]$^{RS}$[CoO$_2$]$_{1.97}$
single crystal. $\rho_{ab}$($T$) shows metallic behavior down to 2
K. This contrasts to the semiconducting behavior below 80 K in the
undoped polycrystalline sample.\cite{Maignan3} It suggests that  K
doping on the Ba sites induces excess charge carriers into the
system. Figure 2c indicates that the metallic in-plane resistivity
below 30 K follows $T^2$-dependence, indicative of strong electron
correlation in the crystal. Such behavior is similar to that
observed in Na$_{0.7}$CoO$_{2}$.\cite{LiSY} However, out-of-plane
resistivity shows a weak semiconducting behavior below 12 K.
Figure 2d indicates that the weak semiconducting behavior obeys
thermal activation behavior with an activation energy of 0.013
meV. At high temperature, the out-of-plane resistivity shows
metallic behavior up to room temperature. No
incoherent-to-coherent behavior is observed in this crystal,
contrasting to the Pb-doped Bi$_2$Ba$_2$Co$_2$O$_y$,\cite{Valla}
in which $\rho_{ab}$($T$) is metallic down to 2 K, while
$\rho_c$($T$) shows a incoherent-to-coherent transition at about
180 K with decreasing temperature. It should be pointed out that
metallic behavior in $\rho_{ab}$($T$) and semiconducting-like
behavior in $\rho_c$($T$) is similar to the case of underdoped
sample in the high-$T_c$ cuprates, in which $\rho_{ab}$($T$) shows
metallic behavior, while $\rho_c$($T$) becomes divergency at low
temperature.\cite{chenxh} This behavior is ascribed to opening of
pseudogap. However, similar behavior in
[Bi$_2$Ba$_{1.3}$K$_{0.6}$Co$_{0.1}$]$^{RS}$[CoO$_2$]$_{1.97}$
single crystal cannot be understood so far.  Figure 2b shows the
temperature dependence of anisotropy $\rho_c/\rho_{ab}$. The
$\rho_c/\rho_{ab}$ is larger than $\sim$10$^3$ in the whole
temperature, suggesting the highly anisotropic electronic
structure. The anisotropy $\rho_c/\rho_{ab}$ increases with
decreasing temperature, similar to that observed in the
(Bi,Pb)$_2$Sr$_2$Co$_2$O$_y$ crystals.\cite{Luo}  Strongly
temperature-dependent anisotropy indicates that scattering
mechanism is different between in-plane and along c-axis,
suggesting a quasi two-dimensional electronic
structure.\cite{chenxh}

Figure 3 shows the temperature dependence of the Hall coefficient
($R_H$) of the crystal. $R_H$ is positive and shows strongly
temperature dependent. Charge carrier concentration can be
estimated from $R_H$ at the room temperature to be
1.33$\times$10$^{21}$ cm$^{-3}$, three times less than that
observed in Na$_{0.7}$CoO$_2$ \cite{WangYY} and two times larger
than that observed in Ca$_3$Co$_4$O$_9$.\cite{Luo1} In the inset
of Fig. 3, it is found that the high-temperature $R_H$ is
proportional to $T^2$, different from the linear temperature
dependent $R_H$ in Na$_{0.7}$CoO$_{2}$.\cite{WangYY} This suggests
the different electronic structure between present misfit-layered
cobaltite and Na$_x$CoO$_2$. Although $\rho_{ab}$ is metallic down
to 2 K, the $R_H$ shows an upturn below about 70 K with decreasing
temperature. Similar behavior in Hall coefficient has been
observed in (Bi,Pb)$_2$Sr$_2$Co$_2$O$_y$ single
crystals,\cite{Yamamoto} in which the upturn of $R_H$ at low
temperature is ascribed to anomalous Hall effect. However, the
$\rho_{xy}$ is linear to magnetic field down to 10 K in the
present crystal, the anomalous Hall effect is not applicable here
to interpret the upturn of $R_H$. The upturn of $R_H$ may suggest
either some localization of charge carrier at low temperature in
spite of the metallic $\rho_{ab}(T)$ down to 2 K or reduction of
density of states at Fermi surface because of some unknown
reasons.

Figure 4 shows the temperature dependence of the thermoelectric
power (TEP). The value of TEP at room temperature is
$\approx$+110$\mu$V/K, as large as that in Na$_{0.75}$CoO$_2$.
However,
[Bi$_2$Ba$_{1.3}$K$_{0.6}$Co$_{0.1}$]$^{RS}$[CoO$_2$]$_{1.97}$ has
larger resistivity than that of Na$_{0.75}$CoO$_2$.\cite{Cavax}
TEP decreases with decreasing temperature. The calculated power
factor $Q$ = $S^2$/$\rho$ is 3.2$\times$10$^{-4}$W m$^{-1}$
K$^{-2}$, larger than those observed in Bi$_2$Ca$_2$Co$_2$O$_y$
($\approx$ 2.7$\times$10$^{-4}$W m$^{-1}$ K$^{-2}$) and
Ca$_3$Co$_4$O$_9$ ($\approx$ 1.8$\times$10$^{-4}$W m$^{-1}$
K$^{-2}$) single crystals.\cite{Luo2} $Q$ increases to
6.5$\times$10$^{-4}$W m$^{-1}$ K$^{-2}$ at around 100 K and then
decreases with decreasing temperature. This maximum is much larger
than those obtained maximum of $Q$ in Bi$_2$Ca$_2$Co$_2$O$_y$
($\approx$ 3.2$\times$10$^{-4}$W m$^{-1}$ K$^{-2}$) and
Ca$_3$Co$_4$O$_9$ ($\approx$ 2.1$\times$10$^{-4}$W m$^{-1}$
K$^{-2}$) single crystals.\cite{Luo2} This indicates that
[Bi$_2$Ba$_{1.3}$K$_{0.6}$Co$_{0.1}$]$^{RS}$[CoO$_2$]$_{1.97}$
single crystal has a better thermoelectric performance than those
in Bi$_2$Ca$_2$Co$_2$O$_y$ and Ca$_3$Co$_4$O$_9$ single
crystals.\cite{Luo2} But the thermoelectric performance of
[Bi$_2$Ba$_{1.3}$K$_{0.6}$Co$_{0.1}$]$^{RS}$[CoO$_2$]$_{1.97}$
single crystal is lower than those observed in Na$_x$CoO$_2$
($x\geq$0.70) according to the data reported by Lee et
al.\cite{Cavax}

\begin{figure}[htbp]
\includegraphics[width=0.48\textwidth]{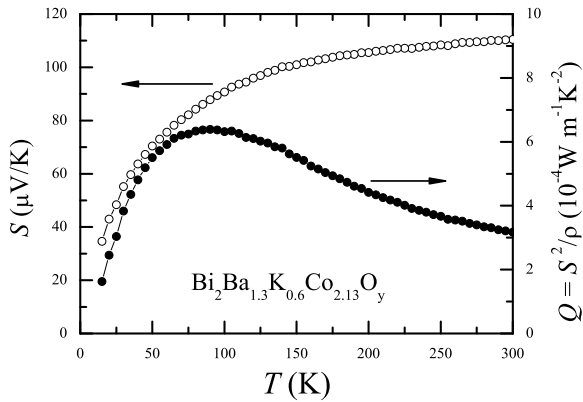}
\caption{Temperature dependence of thermoelectric power (TEP) and
power factor for
[Bi$_2$Ba$_{1.3}$K$_{0.6}$Co$_{0.1}$]$^{RS}$[CoO$_2$]$_{1.97}$
single crystal.}
\end{figure}

\begin{figure}[htbp]
\includegraphics[width=0.5\textwidth]{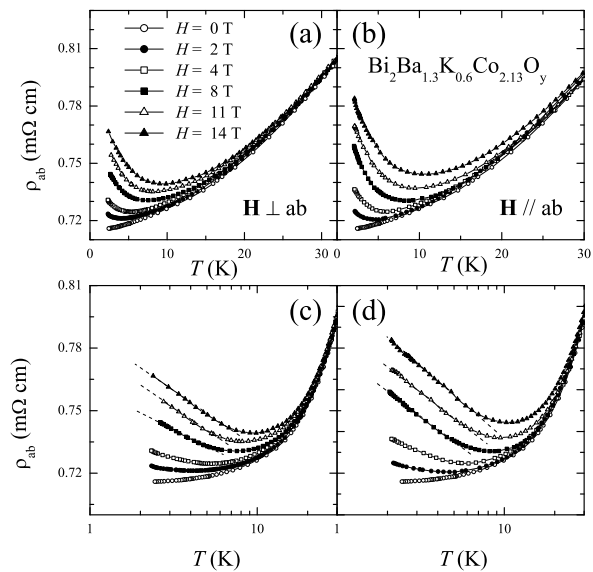}
\caption{Temperature dependence of in-plane magnetoresistivity
under different magnetic field (a): perpendicular and (b):
parallel to $ab$ plane for the
[Bi$_2$Ba$_{1.3}$K$_{0.6}$Co$_{0.1}$]$^{RS}$[CoO$_2$]$_{1.97}$
single crystal. (c) and (d) are replotted of (a) and (b) in
log($T$) scale, respectively. The dashed lines guide eyes linear
to ln(1/$T$).}
\end{figure}

\subsection{Magnetotransport}

Figure 5 shows the temperature dependence of in-plane resistivity
under different magnetic fields ($H$) varying from 0 to 14 T.
Striking feature is observed that magnetic field leads to a
transition from metallic to semiconductor-like behavior at low
temperature with $H$ either perpendicular or parallel to the $ab$
plane. The minimum resistivity appears in $\rho_{ab}(T)$ and
obvious positive magnetoresistance (MR) is observed. As shown in
Fig.5, the temperature corresponding to the minimum resistivity
shifts to high temperature with increasing magnetic field. At 2.5
K and 8 T, the positive MR reaches 4$\%$ as $H$ lies in $ab$ plane
and 6$\%$ as $H$ is along the c-axis, respectively. This value is
much lower than that observed in undoped polycrystalline sample,
where about 10$\%$ of MR was observed at 2.5 and 7
T.\cite{Maignan3} The effect of $H$ on $\rho_{ab}(T)$ is stronger
with $H//ab$ plane than that with $H\perp ab$ plane. As shown in
Fig.6, similar effect of H on $\rho_c(T)$ is also observed.
Although the upturn of $\rho_c(T)$ at low temperature follows a
thermal activation behavior without magnetic field, external
magnetic field leads to a change of low temperature resistivity
from a thermal activation behavior to a ln(1/T) diverging behavior
with $H$ either perpendicular or parallel to the $ab$ plane. It
suggests that $\rho_{ab}(T)$ and $\rho_c(T)$ show the same
temperature dependence under H at low temperature although they
show contrasting behavior (metallic in $\rho_{ab}(T)$ and
semiconducting in $\rho_c(T)$) without H.

\begin{figure}[htbp]
\includegraphics[width=0.48\textwidth]{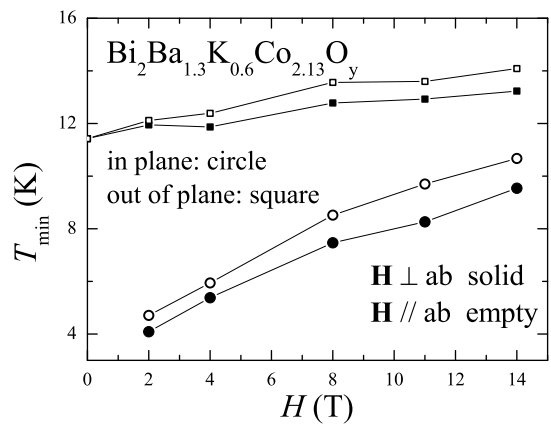}
\caption{Temperature dependence of out-of-plane magnetoresistivity
under different magnetic field (a): perpendicular and (b):
parallel to $ab$ plane for the
[Bi$_2$Ba$_{1.3}$K$_{0.6}$Co$_{0.1}$]$^{RS}$[CoO$_2$]$_{1.97}$
single crystal.}
\end{figure}

Figure 7 shows the magnetic field dependence of the temperature
corresponding to the minimum resistivity ($\rho_{ab}(T)$ and
$\rho_c(T)$). It indicates that the temperature corresponding to
the minimum resistivity ($T_{min}$) increases with increasing H,
suggesting that the localization is enhanced by $H$. It should be
pointed out that $T_{min}$ in $\rho_c(T)$ is enhanced slightly
with increasing magnetic field, contrasting to the strong
dependence of $T_{min}$ in $\rho_{ab}(T)$. Negative MR is a common
feature in Ca$_3$Co$_4$O$_9$ and (Bi,Pb)$_2$$M_2$Co$_2$O$_y$($M$ =
Sr and Ca), in which semiconducting resistivity can usually be
observed below a certain
temperature.\cite{Maignan,Yamamoto,Maignan2} While with $M$ = Ba,
large positive MR has been observed in polycrystalline
sample.\cite{Maignan3} The negative MR has been explained to be
related to the spin-dependent transport at temperatures below or
close to the magnetic ordering transitions, while the positive MR
is not clearly understood. In the previous reports, no such
obvious positive MR has been found in a $\emph{metallic}$
triangular cobaltites: either Tl$_{0.4}$SrCoO$_x$ or
Na$_x$CoO$_2$.  Na$_{0.75}$CoO$_2$ is exceptional case, spin
density wave (SDW) has been taken into account for interpreting
the anomalously large positive MR in it.\cite{Motohashi} We tried
to fit the $\rho_{ab}(T)$ at $H$=8, 11 and 14 T below $T_{min}$
with a variety of functional forms. It turned out that the
resistivity at low temperature under H cannot be fitted by the
formula including thermal activation (ln$\rho\sim$ -1/$T$),
various types of variable range hopping (VRH) conduction
(ln$\rho\sim$ -$T^{-\alpha}$ with $\alpha$=$\frac{1}{2}$,
$\frac{1}{3}$ and $\frac{1}{4}$) and power law (ln$\rho\sim$
ln$T$). Instead, the data at high magnetic field with $H$ either
perpendicular or parallel to the $ab$ plane exhibit a ln(1/T)
divergence [$\rho\sim$ln(1/T)] as shown in Fig.5c, 5d and Fig.6.
Such magnetic-field-induced localization has not been observed yet
in the triangular layered cobaltites previously.

\begin{figure}[htbp]
\includegraphics[width=0.45\textwidth]{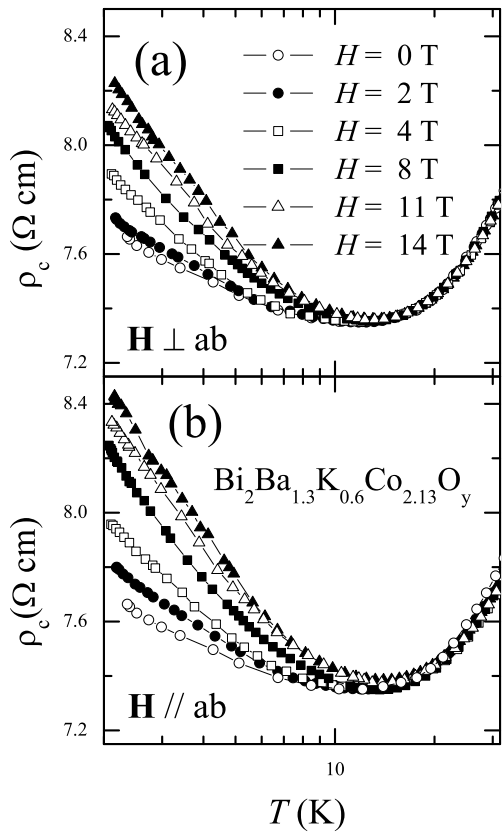}
\caption{The temperature corresponding to the minimum
$\rho_{ab}(T)$ and $\rho_{c}(T)$ as a function of magnetic field
with H parallel and perpendicular to $ab$ plane for the
[Bi$_2$Ba$_{1.3}$K$_{0.6}$Co$_{0.1}$]$^{RS}$[CoO$_2$]$_{1.97}$
single crystal. }
\end{figure}

\begin{figure}[htbp]
\includegraphics[width=0.48\textwidth]{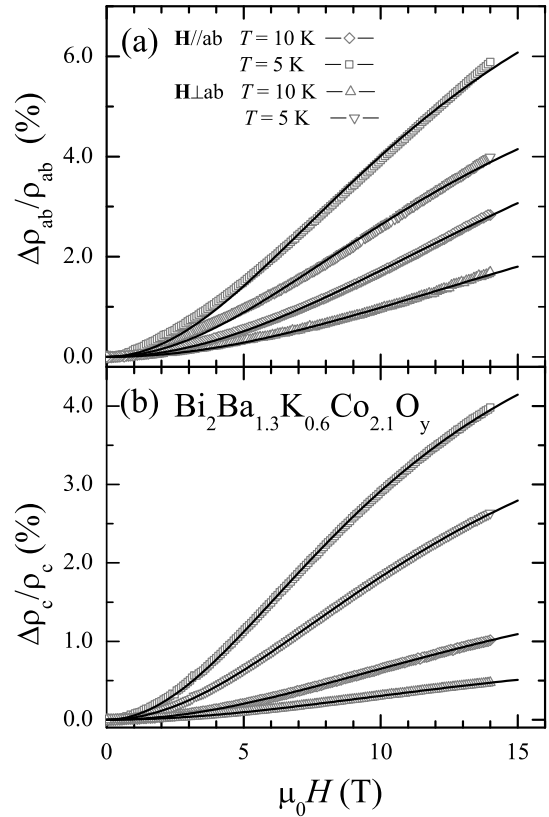}
\caption{Magnetic-field dependence of the in-plane and
out-of-plane isothermal MR with H parallel and perpendicular to
$ab$ plane for the
[Bi$_2$Ba$_{1.3}$K$_{0.6}$Co$_{0.1}$]$^{RS}$[CoO$_2$]$_{1.97}$
single crystal. The solid lines are the fitting results with
Eqn.(1).}
\end{figure}

The in-plane and out-of-plane isothermal MRs were measured  at 5 K
and 10 K with $H$ varying from 0 to 14 T as shown in Fig. 8. All
the MRs are positive and increase with lowering temperature. The
value of MR with $H//ab$ is larger than that with $H//c$ at a
fixed temperature. The MR does not follow the classical $H^2$ law
or Kohler's law (i.e. collapsing to a single curve in Kohler plot,
$\delta\rho/\rho$ vs. $H/\rho_0$). The MRs do not follow a linear
relationship to $H$, as claimed by Hervieu et al. in undoped
polycrystalline sample.\cite{Maignan3} These unusual positive MRs
may be related to the magnetic-field-induced localization as
indicated in Fig.5 and Fig.6.

\section{DISCUSSIONS and CONCLUSIONS}
Magnetotransport behavior can usually be related to the Fermi
surface.\cite{Pippard} For example, a square Fermi surface can lead
to a linear MR due to the presence of a sharp corners.\cite{Pippard}
For systems with multiband electronic structure involving two types
of charge carriers, the $H$ dependence of MR can be written as
\begin{eqnarray}
\Delta\rho/\rho(0)=aH^2/(b+cH^2) \end{eqnarray} where $a$, $b$ and
$c$ are positive, $H$-independent quantities determined by the
relaxation rates of each type of charge carrier.\cite{Ziman} Local
density approximation (LDA) calculations in NaCo$_2$O$_4$
predicted a large cylindrical Fermi surface around $\Gamma$-$A$
line, surrounding by small satellite hole-like pockets centered
about 2/3 of the way out on the $\Gamma$-$K$ and $A$-$H$
directions.\cite{Singh} In Na$_x$CoO$_2$, however, no such
satellite hole-like pocket has been experimentally
observed.\cite{Hasan} Interestingly, Feng et al. found that the
square BiO layer is also conducting except for the triangle Co-O
layer in the same
[Bi$_2$Ba$_{1.3}$K$_{0.6}$Co$_{0.1}$]$^{RS}$[CoO$_2$]$_{1.97}$
single crystal through ARPES.\cite{Feng} The electronic structure
of [Bi$_2$Ba$_{1.3}$K$_{0.6}$Co$_{0.1}$]$^{RS}$[CoO$_2$]$_{1.97}$
is different from that of Na$_x$CoO$_2$. Therefore, multiband
electronic structure does exist in
[Bi$_2$Ba$_{1.3}$K$_{0.6}$Co$_{0.1}$]$^{RS}$[CoO$_2$]$_{1.97}$
single crystal. Furthermore, because of the peculiar rhombohedral
coordination the $t_{2g}$ orbitals of the low-spin Co$^{3+}$ and
Co$^{4+}$ in the triangular cobaltites would be split, forming
heavy $a_{1g}$ and light $e'_{g}$ holes.\cite{Singh,Mizokawa}
Consequently, we tried to fit the MR data shown in Fig.8 using
Eqn.(1). Surprisingly, the out-of-plane MR data can be well fitted
using Eqn.(1) as shown in Fig.8. The in-plane MR data can also be
roughly fitted. Therefore, it suggests that the positive MR in
this crystal possibly results from the multiband electronic
structure with two types of charge carriers.

Slight deviation from Eqn.(1) is observed for the in-plane
isothermal MRs. Therefore, there could be some other mechanisms
for this positive MR in addition to the multiband electronic
structure. A magnetic-field-induced logarithmic temperature
dependence of in-plane resistivity has been observed as shown in
Fig. 5c and 5d. Such magnetic-field-induced logarithmic
temperature-dependent resistivity has been reported in high-$T_c$
cuprates, in which the logarithmically temperature-dependent
resistivity appears when superconductivity is suppressed by
external magnetic field.\cite{Ando} There are two main points of
view to interpret such logarithmic resistivity. In cuprates,
\emph{field-induced} SDW state has been taken as one of the points
of view to interpret the logarithmically temperature-dependent
resistivity.\cite{SunXF,Komiya,Luo3} In most misfit-layered
triangular cobaltitles, such as Ca$_3$Co$_4$O$_9$,
(Bi,Pb)$_2$M$_2$Co$_2$O$_y$ ($M$=Ca and Sr), SDW is a common spin
ordered state.\cite{Sugiyama} Determining from the $\mu$SR
results, however, no SDW exists down to 1.8 K at zero or low field
in updoped Bi$_2$Ba$_2$Co$_2$O$_y$,\cite{Sugiyama} with large
positive MR at low temperature. It could be believed that
[Bi$_2$Ba$_{1.3}$K$_{0.6}$Co$_{0.1}$]$^{RS}$[CoO$_2$]$_{1.97}$
single crystal have no SDW state because of the much better
metallicity compared to the updoped Bi$_2$Ba$_2$Co$_2$O$_y$
sample. If the logarithmically temperature-dependent resistivity
in [Bi$_2$Ba$_{1.3}$K$_{0.6}$Co$_{0.1}$]$^{RS}$[CoO$_2$]$_{1.97}$
single crystal arise from the SDW order as that in high-$T_c$
cuprates, the SDW state should be field-induced. Up to now,
however, no evidence for the field-induced SDW has been reported.
Consequently, $\mu$SR experiments in high magnetic field could be
desired to turn out weather the SDW can be induced in the present
compound by magnetic field.

Another point of view is the weak localization in 2D
systems,\cite{Hidaka,Hagen} which theoretically predicts a
logarithmic temperature dependence of the conductivity.\cite{log}
The misfit structure leads to the system to be 2-dimensional
structurally. The large anisotropy and the strong temperature
dependence of the anisotropy as shown in Fig.2 suggests the
[Bi$_2$Ba$_{1.3}$K$_{0.6}$Co$_{0.1}$]$^{RS}$[CoO$_2$]$_{1.97}$
single crystal to be a (quasi) 2D electronic system. Evidencing
from these results, it is also possible that field-induced 2D weak
localization results in the logarithmically temperature-dependent
resistivity. This is reason why the in-plane MR data are roughly
fitted with slight deviation based on the multiband electronic
structure with two types of charge carriers.

 In conclusion, resistivity under magnetic field, thermopower and Hall coefficient
are systematically studied for
[Bi$_2$Ba$_{1.3}$K$_{0.6}$Co$_{0.1}$]$^{RS}$[CoO$_2$]$_{1.97}$
single crystal. An anomalous behavior is observed that there exist
a contrasting behavior at low temperature in in-plane resistivity
($\rho_{ab}$(T)) and out-of-plane ($\rho_{c}(T)$):  metallic
behavior down to 2 K with a $T^2$ dependence below 30 K in
$\rho_{ab}$(T); while a thermal activation semiconducting behavior
below about 12 K in $\rho_{c}(T)$. Magnetic field leads to a
ln(1/$T$) diverging behavior in both $\rho_{ab}$ and $\rho_{c}(T)$
at low temperature. The isothermal out-of-plane MR can be quite
well fitted by taking into account the multiband electronic
structure with two types of charge carriers. The ln(1/$T$)
diverging $\rho_{ab}(T)$ in magnetic field could arise from the
field-induced 2D weak localization or magnetic field induced spin
density wave.

\section{ACKNOWLEDGEMENT}
This work is supported by the National Natural Science Foundation
of China and by the Ministry of Science and Technology of China
(973 project No: 2006CB601001 and
2006CB0L1205).\\

\vspace*{5mm} \noindent
 $^{\ast}$ Corresponding author. \emph{Electronic
address:} chenxh@ustc.edu.cn


\begin{thebibliography}{}

\bibitem{Teresaka}
I. Terasaki, Y. Sasago, and K. Uchinokura, Phys. Rev. B \textbf{56},
R12685 (1997).

\bibitem{Maignan}
A. C. Masset, C. Michel, A. Maignan, M. Hervieu, O. Toulemonde, F.
Studer, and B. Raveau, Phys. Rev. B \textbf{62}, 166 (2000).

\bibitem{WangYY}
Y. Y. Wang, N. S. Rogado, R. J. Cava, and N. P. Ong, (London)
\textbf{423}, 425(2003); Y. Y. Wang, N. S. Rogado, R. J. Cava, and
N. P. Ong, cond-mat/0305455.

\bibitem{LiSY}
S. Y. Li, L. Taillefer, D. G. Hawthorn, M. A. Tanatar, J. Paglione,
M. Sutherland, R. W. Hill, C. H. Wang, and X. H. Chen, Phys. Rev.
Lett. \textbf{93}, 056401 (2004).

\bibitem{Takada}
K. Takada, H. Sakurai, E. Takayama-Muromachi, F. Izumi, R. A.
Dilanian, and T. Sasaki, Nature \textbf{422}, 53 (2003).

\bibitem{Yamamoto}
T. Yamamoto, I. Tsukada, M. Takagi, T. Tsubone, and K. Uchinokura,
J. Magn. Magn. Mater. \textbf{226-230}, 2031 (2001); T. Yamamoto, K.
Uchinokura, and I. Tsukada, Phys. Rev. B \textbf{65}, 184434 (2002).

\bibitem{Maignan2}
A. Maignan, S. Hebert, M. Hervieu, C. Machel, D. Pelloquin and D.
Khomskii, J. Phys: Condens. Matter \textbf{15}, 2711 (2003).

\bibitem{Cava}
M. L. Foo, Y. Y. Wang, S. Watauchi, H. W. Zandbergen, T. He, R.J.
Cava, and N.P. Ong, Phys. Rev. Lett. \textbf{92}, 247001 (2004).

\bibitem{WangCH}
C. H. Wang, X. H. Chen, T. Wu, X. G. Luo, G. Y. Wang, and  J. L. Luo
Phys. Rev. Lett. \textbf{96}, 216401(2006).

\bibitem{Sugiyama}
J. Sugiyama, J. H. Brewer, E. J. Ansaldo, H. Itahara, T. Tani, M.
Mikami, Y. Mori, T. Sasaki, S. Hebert, and A. Maignan, Phys. Rev.
Lett. \textbf{92}, 017602 (2004).

\bibitem{Leigylin}
H. Leligny, D. Grebille, O. Perez, A. C. Masset, M. Hervieu, C.
Michel, and B. Raveau, C. R. Sci. Paris IIc, Chim \textbf{2}, 409
(1999); Acta Crystallogr., Sect. B: Struct. Sci. \textbf{56}, 173
(2000).

\bibitem{Maignan3}
M. Hervieu, A. Maignan, C. Michel, V. Hardy, N. Creon, and B.
Raveau, Phys. Rev. B \textbf{67}, 045112 (2003).

\bibitem{Luo}
X. G. Luo, X. H. Chen, G. Y. Wang, C. H. Wang, X. Li, W. J. Miao, G.
Wu, and Y. M. Xiong, Eur. Phys. J. B. \textbf{49},37 (2006).

\bibitem{Shannon}
R. D. Shannon, Acta Crystallogr. \textbf{A32}, 751 (1976); R. D.
Shannon, C. T. Prewitt, Acta Crystallogr. \textbf{B25}, 925 (1969).

\bibitem{Valla}
T. Valla, P. D. Johnson, Z. Yusof, B. Wells, Q. Li, S. M. Loureiro,
R. J. Cava, M. Mikami, Y. Mori, M. Yoshimura, and T. Sasaki, Nature
\textbf{417}, 627 (2002); Z. Yusof, B.O. Wells, T. Valla, P.D.
Johnson, A.V. Fedorov, Q. Li, S.M. Loureiro, R.J. Cava,
Cond-mat/0610271.

\bibitem{chenxh}
X. H. Chen, M. Yu, K. Q. Ruan, S. Y. Li, Z. Gui, G. C. Zhang, and
L. Z. Cao, Phys. Rev. B {\bf 58}, 14219(1998).


\bibitem{Luo1}
X. G. Luo, X. H. Chen, C. H. Wang, G. Y. Wang, Y. M. Xiong, H. B.
Song, H. Li, and X. X. Lu Europhys. Lett. \textbf{74}, 526 (2006).

\bibitem{Cavax}
M. Lee, L. Viciu, Lu Li, Y. Y. Wang, M. L. Foo, S. Watauchi, R. A.
Pascal Jr., R. J. Cava, and N. P. Ong, Nature Materials
\textbf{425}, 537 (2006).

\bibitem{Luo2}
X. G. Luo, Y. C. Jing, H. Chen, and X. H. Chen, J. Cryst. Growth, in
press.

\bibitem{Motohashi}
T. Motohashi, R. Ueda, E. Naujalis, T. Tojo, I. Terasaki, T. Atake,
M. Karppinen, and H. Yamauchi, Phys. Rev. B \textbf{67}, 064406
(2003).

\bibitem{Pippard}
A. B. Pippard, \textsl{Magnetoresistance in Metal} (Cambridge
University Press, Cambridge, 1989).

\bibitem{Ziman}
J. M. Ziman, \textsl{Principles of the Theory of Solids}, 2nd ed.
(Cambridge University Press, Cambridge, 1972).

\bibitem{Singh}
D. J. Singh, Phys. Rev. B \textbf{61}, 13397 (2000).

\bibitem{Hasan}
M. Z. Hasan, Y. D. Chuang, D. Qian, Y.W. Li, Y. Kong, A. Kuprin,
A.V. Fedorov, R. Kimmerling, E. Rotenberg, K. Rossnagel, Z. Hussain,
H. Koh, N.S. Rogado, M.L. Foo, and R.J. Cava, Phys. Rev. Lett.
\textbf{92}, 246402 (2004).

\bibitem{Feng}
D. L. Feng, private communication.

\bibitem{Mizokawa}
T. Mizokawa, L. H. Tjung, P. G. Steeneker, N. B. Brookes, I.Tsukada,
T. Yamamoto, and K. Uchinokura, Phys. Rev. B \textbf{64}, 115104
(2001).

\bibitem{Ando}
Y. Ando, G. S. Boebinger, A. Passner, T. Kimura, and K. Kishio,
Phys. Rev. Lett. \textbf{75}, 4662 (1995).

\bibitem{SunXF}
X. F. Sun, S. Komiya, J. Takeya, and Y. Ando, Phys. Rev. Lett.
\textbf{90} 117004 (2003).

\bibitem{Komiya}
S. Komiya and Y. Ando, Phys. Rev. B \textbf{70} 060503 (2004).

\bibitem{Luo3}
X. G. Luo, X. H. Chen, X. Liu, R. T. Wang, C. H. Wang, L. Huang, L.
Wang, and Y. M. Xiong, Supercond. Sci. $\&$ Technol. \textbf{18} 234
(2005).

\bibitem{Hidaka}
Y. Hidaka, Y. Yamaji, K. Sugiyama, F. Tomiyama, A. Yamagishi, M.
Date, M. Hikita, J. Phys. Soc. Jpn. \textbf{60}, 1185 (1991).

\bibitem{Hagen}
S. J. Hagen, X. Q. Xu, W. Jiang, J. L. Peng, Z. Y. Li and R. L.
Greene, Phys. Rev. B \textbf{45}, 515 (1992).

\bibitem{log}
G. Bergmann, Phys. Reports \textbf{107}, 1 (1984).

\end{thebibliography}
\end{document}